\documentclass[aps,prl,twocolumn,letterpaper,showpacs]{revtex4-1}
\usepackage[utf8x]{inputenc}
\usepackage{graphicx,amsmath,mathrsfs}
\usepackage{amssymb}
\usepackage{amsthm}
\usepackage{bm}

\def\bc{\begin{center}}
\def\ec{\end{center}}

\newcommand{\ket}[1]{\left|#1\right\rangle}

\renewcommand {\ec}{\eta_{\gamma}}

\newcommand{\HH}{{\cal H}}
\newcommand{\GG}{{\cal G}}
\newcommand{\im}{{\mathrm{i}}}

\begin{document}
\title{Tunneling spectra simulation of interacting Majorana wires}
\author{Ronny Thomale${}^{1,2}$}
\author{Stephan Rachel${}^3$}
\author{Peter Schmitteckert${}^4$}
\affiliation{${}^{1}$Institut de th\'eorie des ph\'enom\`enes
  physiques, \'Ecole Polytechnique F\'ed\'erale de Lausanne (EPFL),
  CH-1015 Lausanne}
\affiliation{${}^{2}$Theoretical Physics, University of W\"urzburg, D-97074
  W\"urzburg, Germany} 
\affiliation{${}^3$Institute for Theoretical Physics, Dresden University of Technology, 01062 Dresden, Germany}
\affiliation{${}^4$Institut f\"ur Nanotechnologie, Forschungszentrum
  Karlsruhe, D-76021 Karlsruhe, Germany}

\date{\today}

\begin{abstract}
  Recent tunneling experiments on InSb hybrid
  superconductor-semiconductor devices have provided hope for a
  stabilization of Majorana edge modes in a spin-orbit
  quantum wire subject to a magnetic field and superconducting
  proximity effect. Connecting the experimental scenario with a
  microscopic description poses challenges of different kind, such
  as accounting for the effect of interactions on the tunneling
  properties of the wire. We develop a density matrix renormalization
  group (DMRG) analysis of the tunneling spectra of interacting
  Majorana chains, which we explicate for the Kitaev chain
  model. Our DMRG approach allows us to calculate the spectral
  function down to zero frequency, where we analyze how the Majorana
  zero-bias peak is affected by interactions. From the study of
  topological phase transitions between the topological and trivial superconducting
  phase in the wire, we argue that the bulk gap closure generically affects both
  the proximity peaks and the Majorana peak, which should be
  observable in the transport signal.
\end{abstract}


\maketitle

{\it Introduction.} The field of topological phases in correlated
electron systems is witnessing enormous interest in
contemporary condensed matter physics. A new stage has been set by the
field of topological insulators and superconductors, which promoted
the role of spin-orbit coupling from a quantitative relativistic
correction to a substantial system parameter characterizing electronic
quantum states of matter~\cite{hasan-10rmp3045,qi-11rmp1057}.  Aside
from the fundamental significance by its own, this direction
revitalized the search for Majorana bound states (MBS) as soon as Fu and Kane realized that topological
insulators can induce MBS at the surface in proximity to a
superconductor~\cite{fu-08prl096407}, which could be detected through
resonant Andreev tunneling at the
surface~\cite{law-09prl237001}. Along with the challenging
experimental effort to make these interfaces
accessible~\cite{williams-12prl056803}, Sau et al.~\cite{sau-10prl040502}
as well as Alicea~\cite{alicea10prb125318} suggested 
alternative setups for such an effect via composite
compounds of semiconductors and ferromagnetic insulators.  Preceded by
a milestone work of Kitaev~\cite{kitaev01pu131}, this paved the way
for theoretical proposals of one-dimensional versions of this scenario
where a spin-orbit quantum wire is placed in proximity to a
superconductor and subject to an applied magnetic field. There,
Majorana modes are
predicted to appear at the edge of the
wire~\cite{lutchyn-10prl077001,oreg-10prl177002,beenakker13arcmp113,alicea12rpp076501}
and manifest themselves as a conductance
peak~\cite{flensberg10prb180516,law-09prl237001,wimmer-11njp053016}. 
The tunneling
experiments by the Kouwenhoven group~\cite{mourik-12s1222360} along
with subsequent independent accomplishments by other groups employing
tunneling~\cite{nilsson-12nl228,deng-12nl6414,das-12np887, finck-13prl126406} and
Josephson~\cite{rokhinson-12np795} measurements suggest that the
spin-orbit quantum wires are an experimental scenario where MBS might
be detectable: the $\text{InSb}$ wires possess large spin orbit
coupling, and appropriate contacts guarantee high transparency for
electrons to induce superconducting (SC) gaps~\cite{doh-05s272}. At the
same time, the high Land\'e factor of InSb~\cite{nilsson-09nl3151}
assures that one can still efficiently induce spin alignment in the
wire by comparably low magnetic fields which do not significantly
affect the SC proximity effect.

A first microscopic perspective on MBS emerged from the Pfaffian wave
function in the context of 
paired Hall states~\cite{greiter-91prl3205,Moore-91npb362} which was subsequently
connected to the $A$ phase of ${}^3\text{He}$~\cite{volovik-book},
$p+ip$ superconductors~\cite{Read-00prb10267}, and recently to optical
lattice scenarios~\cite{sato-09prl020401} as well as Majorana spin
liquids~\cite{greiter-09prl207203}.
MBS emerge as zero energy midgap states in the vortex solution of the
Bogoliubov-de-Gennes equation~\cite{kopnin-91prb9667,volovik99jetp609,Read-00prb10267,Ivanov01prl268,kitaev01pu131}. 
The
MBS vortex state is protected through the emergent
particle-hole symmetry of the superconductor and exhibits a vortex
energy gap.
Due to lack
of phase space associated with the edges of the wire in the clean
limit, the are no competing midgap states localized at the edge,
suggesting that the MBS are protected by the full proximity gap $\Delta
\sim 1K$~\cite{mourik-12s1222360}. 
Moreover, the tunability of several system parameters should make it feasible to observe the
topological phase transition between a phase with and without
MBS at the edge.

Various effects such as disorder, strength and direction of magnetic field, or temperature have been investigated
for the Majorana
wire~\cite{brouwer-11prl196804,lobos-12prl146403,tewari-12prb024504,lin-12prb224511,stanescu-12prl266402,bagrets-12prl227005,liu-12prl267002,pikulin-12njp125011,pientka-12prl227006,sela-11prb085114,gangadharaiah-11prl036801}. This
is an essential step to further understand experiments, as there are
various alternative resonances induced by Josephson or Andreev bound
states, Kondo physics or disorder-imposed midgap states that could
give rise to similar transport signals. 
Among all of these effects, the role of interactions is most complicated to
address microscopically for a finite wire, as the Hamiltonian looses
its bilinear form. As such, interactions cannot be
easily treated for large system sizes unless a Luttinger liquid
approximation is adopted where the proximity gap can only be included
perturbatively, or interactions can only be considered in special
scaling limits~\cite{lutchyn-11prb214528,fidkowski-12prb245121,lutchyn-arxiv1302} . The mesoscopic limit
$q, \omega\rightarrow 0$ suggests that the low energy
treatment of tunneling experiments only depends on the
existence of Majorana edge modes irrespective of the spectral properties in
the bulk. This assumption, however, is invalid for any AC-type
measurement at finite $\omega$ and for ${\mathrm d}I/{\mathrm d}V_{\rm SD}$
DC measurements at finite bias, where $V_{\rm SD}$ is the source-drain voltage.

In this Letter, we develop a density matrix renormalization group
(DMRG) ansatz to study
the role of interactions in Majorana wires by computing the full spectral function down to
zero frequency. DMRG has been previously
employed to obtain the doubly degenerate ground state of the
Majorana wire~\cite{stoudenmire-11prb014503}. The motivation to formulate a
DMRG ansatz for the full spectral function is two-fold. First, this
allows to investigate the role of interactions on a microscopic level
and connect its effects to the ${\mathrm d}I/{\mathrm d}V_{\rm SD}$ signal. In particular, the
suspected Majorana zero bias peak is centered around zero frequency,
which would be hard to resolve in conventional time-resolved DMRG
where an infinite time evolution would have to be performed. Second, we thus develop
the platform to consider the interplay of effects such as disorder,
temperature, and interactions in a most suited microscopic framework, which is
likely to stimulate a subsequent quantitative
analysis of  experimental scenarios. 

{\it Model.} 
We consider the effective description along the proposal by Kitaev~\cite{kitaev01pu131} for a single
chain and hard wall boundary
conditions:
\begin{eqnarray}
\HH &=&\sum_{i=1}^{M-1} \left( -t c_i^{\dagger}c_{i+1}^{\phantom{\dagger}} +\Delta c_i^{\phantom{\dagger}}
c_{i+1}^{\phantom{\dagger}} + \text{h.c.} \right) -\mu
\sum_{i=1}^M n_i\nonumber \\
&&+\sum_{i=1}^{M-1} V n_i n_{i+1},
\label{model}
\end{eqnarray}
where $n_i=c_i^{\dagger} c_i^{\phantom{\dagger}}$, $M$ denotes the
number of sites, $t$ the nearest neighbor hopping (set to unity in the
following), $\Delta$ the proximity gap, $\mu$ the chemical potential,
and $V$ the nearest neighbor Hubbard interaction. For $V=0$, the
system can be studied analytically in a single-particle
picture~\cite{kitaev01pu131}. As a function of $\mu$, a topological
phase transition is driven between a bulk-gapped SC wire with ($\vert
\mu \vert < 2t$) and without ($\vert \mu \vert \ge 2t$) one Majorana
mode per edge which are still entangled whereas correlations decay at
the scale $\sim$ $1/\Delta$ in the bulk. The spectral signature of
this is given by a ground state degeneracy for the two different
parity sectors $P=(-1)^{\sum_i n_i}$ labeled even ($P=1$) and odd
($P=-1$). For the ground state in the even case, all electrons pair
and avoid the proximity gap scale. For the odd case, the excess
electron pays a Bogoliubov excitation energy $\sim \Delta$ in the
trivial SC phase of the wire, while it can be located in the zero
energy entangled state in the topological SC phase of the wire as
provided by the Majorana edges. Accordingly, a single electron in
transport takes advantage of the zero energy fermionic state formed by
the two Majorana edges, yielding a shift in the quantized conductance
and a zero bias peak in the ${\mathrm d}I/{\mathrm d}V_{\rm SD}$
signal~\cite{flensberg10prb180516,law-09prl237001,wimmer-11njp053016}. In
particular, the energy location of the fermionic state energy formed
by the Majorana edges is protected by particle-hole symmetry: as soon
as the SC phase forms in the wire, the MBS does not evolve in energy
and hence should give a zero bias peak irrespective of modifications
imposed on the wire which leave the specific SC phase intact,
i.e. which do not close the bulk gap. From a different perspective of
one-dimensional systems, the nontrivial phase of~\eqref{model} can
also be labelled topological~\cite{turner-11prb075102} in the sense
that the bulk gap forms without breaking continuous lattice
symmetries, and yields fractionalized edge modes as compared to the
constituent fermions which span the Hilbert space of the system. This
is similar to the Haldane gap scenario of $S=1$ chains where the
featureless bulk is gapped and the edges form $S=1/2$ degrees of
freedom~\cite{haldane83prl1153}.

\begin{figure}[t]
\begin{minipage}{0.99\linewidth}
\includegraphics[width=\linewidth]{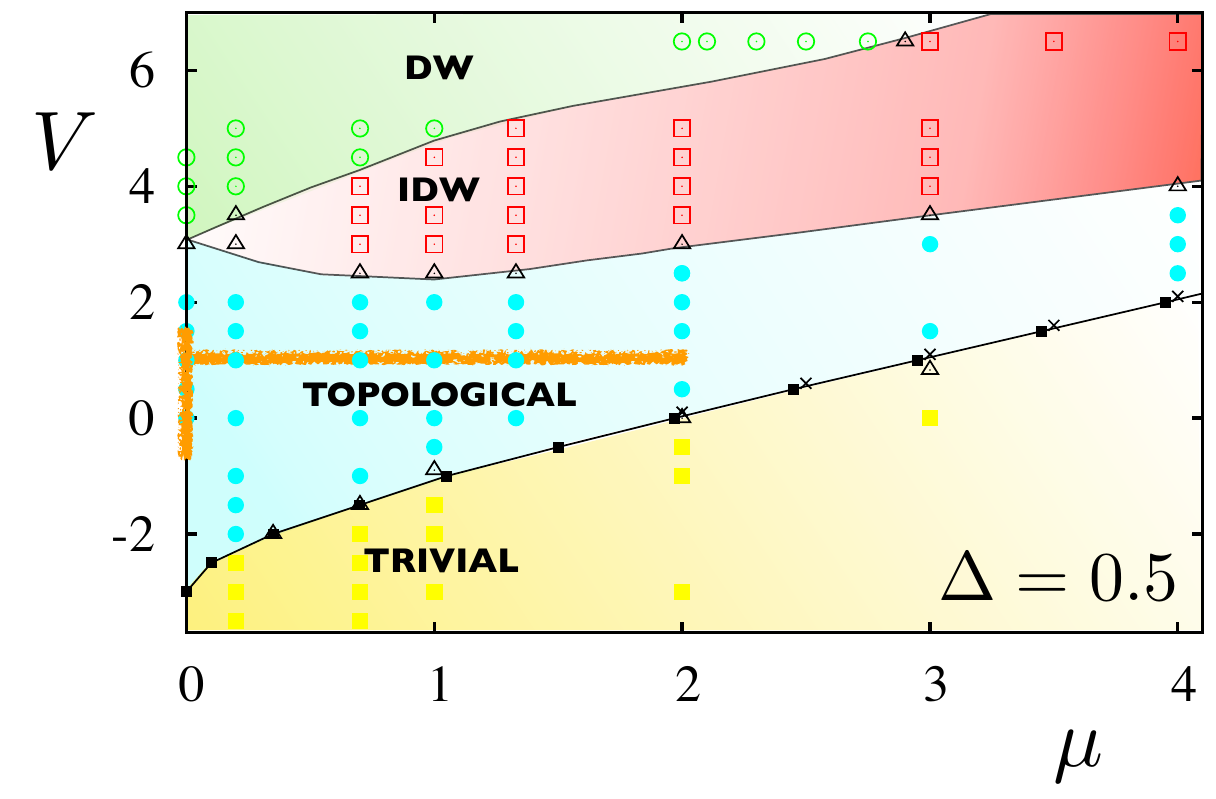}
\end{minipage}
\caption{(Color online). Phase diagram of~\eqref{model}
  for $\Delta=0.5$. Data points are obtained within DMRG for
  different system sizes: A trivial SC
  phase (yellow), the
  topological SC phase (light blue), an incommensurate density
  wave (IDW) (red), and
  an regular density wave (DW) phase (green) is found. Black lines indicate criticality, orange lines
  the parameter regions for Figs.~\ref{fig3} and \ref{fig1}.}
\label{fig4}
\end{figure}

The second line in~\eqref{model} represents the
most short-range interaction term between the fermions allowed by
Pauli principle. While the proximity of the superconductor will be
efficient in screening the long-range part of generic Coulomb
interactions between the electrons, the short-range potential is less affected and needs to be considered. In the
following, we treat finite size realizations of~\eqref{model} up to
$M=200$ for specific points, and compute the spectral function $A(\omega)$, i.e.\ the local 
single particle density of states, which dictates the ${\mathrm d}I/{\mathrm d}V_{\rm SD}$ signal
of a tunneling current $I$.

\begin{figure}[t]
\begin{minipage}{0.99\linewidth}
\includegraphics[width=\linewidth]{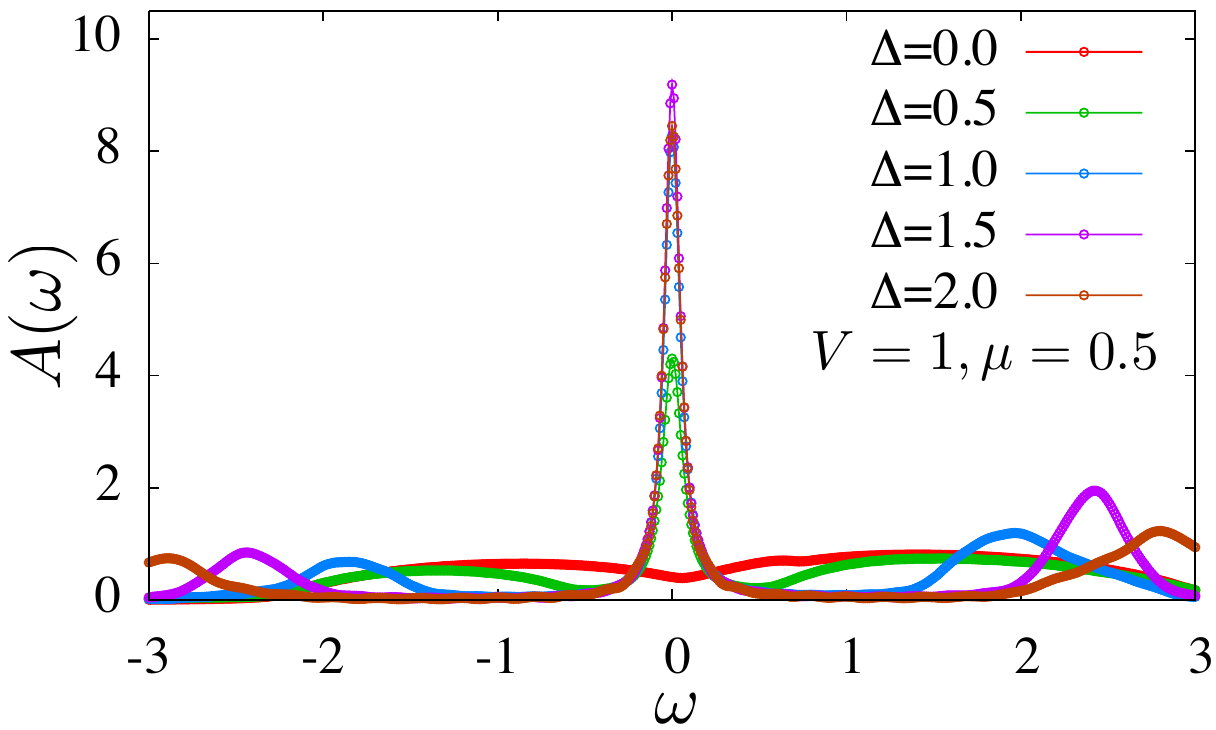}
\end{minipage}
\caption{(Color online). DMRG spectral functions $A(\omega)$ for
  different amplitudes $\Delta$. ($M=96$, $V=1$, and $\mu=0.5$.) 
The proximity peaks are asymmetric due to
  finite $\mu$.}
\label{fig2}
\end{figure}

{\it Method.} The spectral function is obtained from the imaginary part of
the retarded Greens function 
\begin{align}
\GG^{\mathrm r}(z)  &= \GG_{\hat{c}^{}_x, \hat{c}^{+}_x}^+  - \GG_{\hat{c}^{+}_x, \hat{c}^{}_x}^- \\
  \GG^\pm_{\hat{A},\hat{B}}(z) &= \langle \Psi_0 |   \hat{A} \left(
    E_0 -H \pm z\right)^{-1} \hat{B} | \Psi_0 \rangle , \label{eq:Resolvents}
\end{align}
where $\hat{A}$ and $\hat{B}$ are placeholders for the operators of interest ($\hat{c}^{}_{x_0}, \hat{c}^{+}_{x_0}$), $\ket{\Psi_0}$ is the ground state
with energy $E_0$, ${x_0}$ denotes the position where the local density of states is evaluated, and
$z=\omega + \im \eta$ the complex frequency including the level broadening which has to be introduced to smear
over finite size effects~\cite{Schmitteckert:JPCS2010}.
We evaluate the resolvent equations \eqref{eq:Resolvents} by expanding
\begin{equation}
 f_\pm( \HH -E_0, z ) = \frac{1}{ E_0 - \HH \pm z } 
\end{equation}
into Chebyshev orthogonal polynomials $T_n$~\cite{Braun_Schmitteckert:X2012}   
\begin{align}
  f_\pm( z,x  )     &= 1/(\pm z -x)  =  \sum_{n=0}^\infty \alpha^\pm_n(z) T_n(x) \\
  \alpha_\pm(z) &= \frac{2/(1+\delta_{n,0})}{(\pm z)^{n+1}(1+\sqrt{z^2}\sqrt{z^2-1}/z^2)^n \sqrt{1-1/z^2}} .\label{eq:Coefficients}
\end{align}
In contrast to  the standard kernel polynomial scheme~\cite{RevModPhys.78.275}, 
we evaluate the expansion at a finite broadening $\eta$~\cite{Braun_Schmitteckert:X2012,Schmitteckert:JPCS2010}, and the local
density of states is given by
\begin{equation}
  A(\omega) = - \frac{1}{\pi} \lim_{\eta\rightarrow 0^+} \GG^r_{\hat{c},\hat{c}^+}(\omega + \im \eta) .
\end{equation}
The moments $T_n =  \langle \Psi_0 |   T_n\left( E_0 -H \right) | \Psi_0 \rangle $
are obtained using the recurrence relations for the Chebyshev polynomials
and all $ | \zeta_n \rangle = T_n\left( E_0 -H \right) | \Psi_0 \rangle $ states are added to the density matrix
to optimize for the basis at each DMRG step.   
Within the DMRG procedure, we exploit the parity quantum number,
and are typically using at least 1000 states per DMRG block. For calculating the moments for the Chebyshev expansion,
we perform a first calculation for the first ten moments only. We then restart the DMRG to increase the
number of moments in several restarts up to $n=800$.
As for the single-particle limit $V = 0$, we verified our results
against a generalized Bogoliubov transformation~\cite{peterkory}. We
deconvolute the applied $\eta=0.1$ ($0.17$) of the $M=96$ ($48$) site
systems as described in~\cite{Schmitteckert:JPCS2010}.


\begin{figure}[t]
\begin{minipage}{0.99\linewidth}
\includegraphics[width=\linewidth]{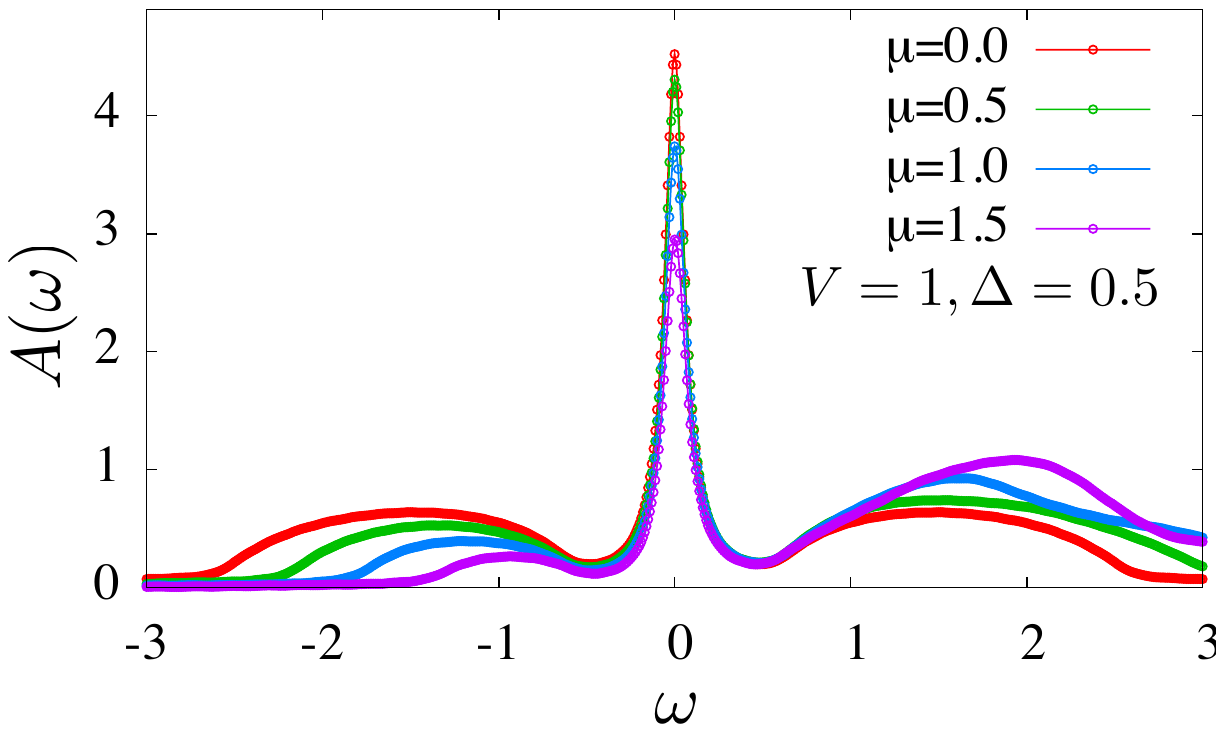}
\end{minipage}
\caption{(Color online). DMRG spectral functions $A(\omega)$ for
  different $\mu$. ($M=96$, $V=1$, and $\Delta=0.5$.)}
\label{fig3}
\end{figure}

{\it $V$-$\mu$ phase diagram.} Fig.~\ref{fig4} displays the numerical
phase diagram as obtained from 
our DMRG
approach: As a function of $V$ and $\mu$, the system can reside in the
trivial and topological SC phase, as well as in an (incommensurate) density
wave state (I)DW for strong repulsive coupling. 
The topological SC phase is detected by the two-fold
degenerate ground states belonging to different parity sectors.
In contrast, the two ground states of
the (I)DW phase belong to the same parity sector, where a distinction
between IDW and DW can be made by analyzing the homogeneity of local
densities and entropy signatures. 
The four different gapped phases are separated from each other by critical lines.
We
observe a strong renormalization of $\mu_c$ separating the trivial and
topological SC phase as a function of interaction strength.
Our numerical phase diagram agrees quite well with the asymptotic
analytic solution obtained by mapping the Kitaev chain to a Josephson junction
array~\cite{dirk}. 

{\it $\Delta$-dependence.} We pick the phase space point $(V,\mu)=(1,0.5)$ located in
the topological SC phase, and enhance the proximity scale $\Delta$. As
soon as $\Delta$ is turned on, we
find a clean Majorana zero bias peak along with proximity peaks around
$\omega = \pm \Delta$. Note that even though the non-interacting
system breaks particle hole symmetry due to finite $\mu$, the spectral
function shows the expected emergent particle-hole symmetry for
$\vert \omega \vert<\Delta$. 

{\it $\mu$-dependence.} We fix $V$ and investigate the behavior of
the spectral function for increasing $\mu>0$ as we trace through the
topological SC phase along the horizontal orange line in Fig.~\ref{fig4}. The hole-like weight gets
increasingly shifted to the electron-like regime, while the Majorana
peak signal is robust independent of $\mu$. 

\begin{figure}
\begin{minipage}{0.99\linewidth}
\includegraphics[width=\linewidth]{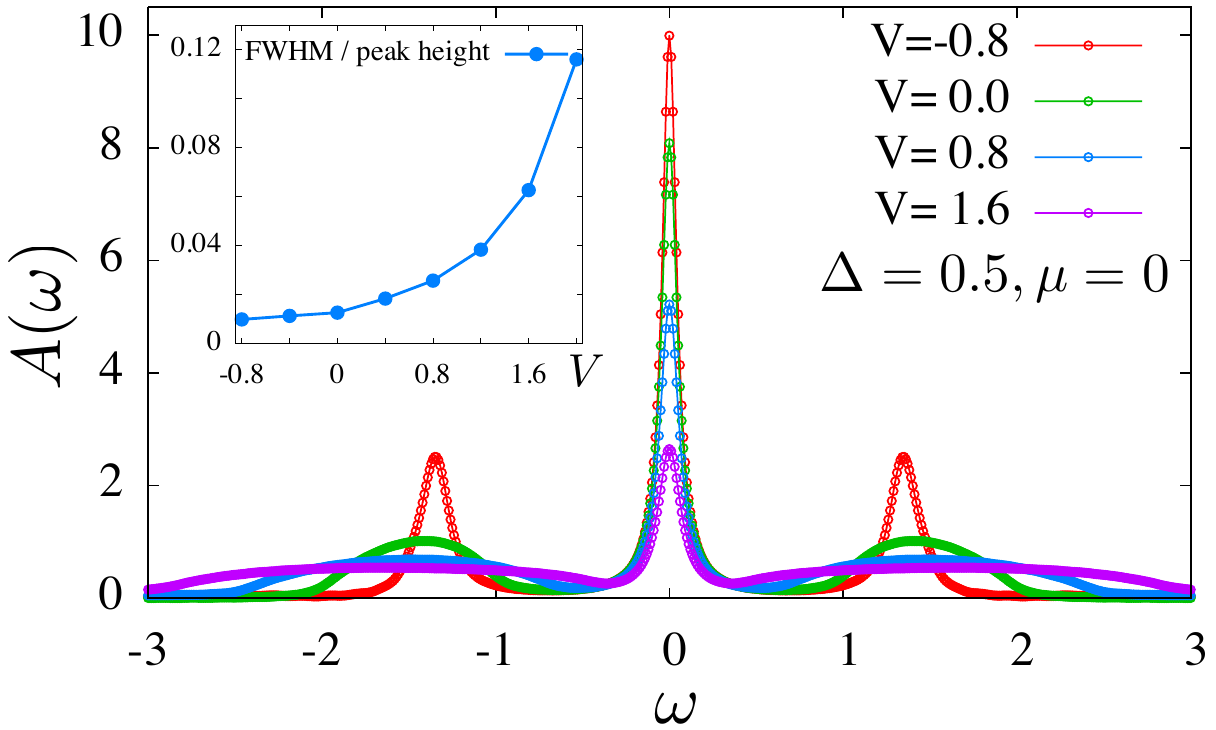}
\end{minipage}
\caption{(Color online). DMRG spectral functions $A(\omega)$ for
  various interaction strengths $V$. ($M=96$, $\Delta=0.5$, and $\mu=0$.) Moderate attractive $V$ increases the Majorana peak
  height while repulsive $V$ suppresses the zero-bias peak. The
  Majorana peak
  broadens as illustrated in
  the inset displaying the FWHM divided by the peak height.}
\label{fig1}
\end{figure}

{\it $V$-dependence.} To show the characteristic behavior of the
spectral function for varying interaction strengths, we trace a
regime of $V$ from weakly attractive to strongly repulsive in the
topological SC regime (Fig.~\ref{fig1}), as depicted by the vertical
orange line in Fig.~\ref{fig4}. Weak attractive $V$ sharpens the
proximity peaks and enhances the Majorana zero bias peak, along with
the effective renormalization of the charge gap. The proximity peaks
become broad due to repulsive $V$. Similarly, the zero bias peak is
sensitive to the interaction strength and quickly decreases in
height as the interactions become repulsive. The inset
in Fig.~\ref{fig1} shows the FWHM divided by peak height of the zero bias peak as a function
of $V$, where a significant broadening is observed. It
suggests that in the actual ${\mathrm d}I/{\mathrm d}V_{\rm SD}$ measurement, the zero bias broadening is
generally a combined effect of temperature and interactions.


\begin{figure}
\begin{minipage}{0.99\linewidth}
\includegraphics[width=\linewidth]{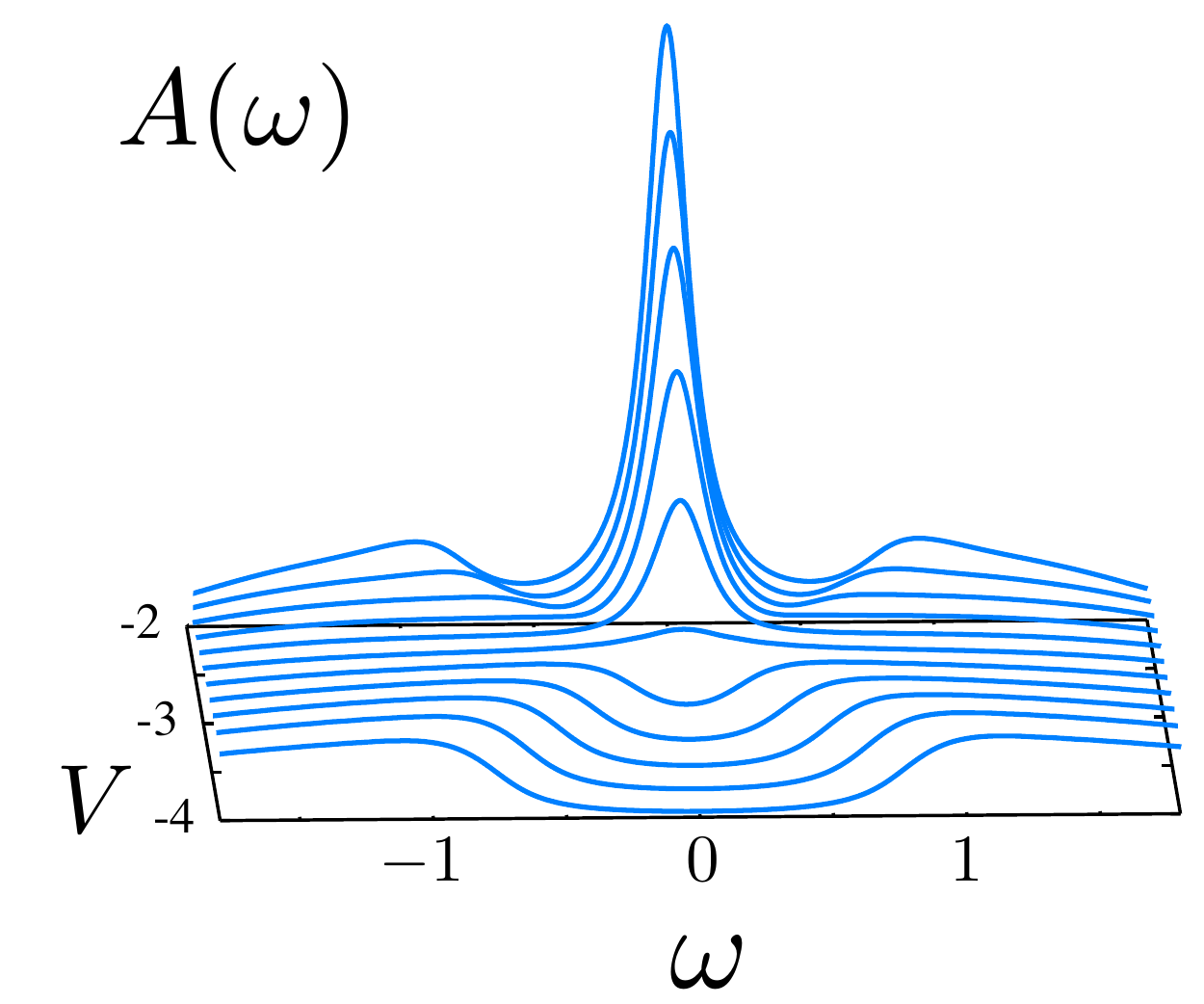}
\end{minipage}
\caption{(Color online). DMRG spectral functions $A(\omega)$ in the
  regime $V=-2,\dots,-4$ in increments of 0.2. ($M=48$, $\Delta=0.5$, and $\mu=0$.) The bulk gap closure induces a joint
  collapse of the Majorana peak and the proximity gaps until the
  latter reopen in the trivial superconducting phase (Fig.~\ref{fig4}).}
\label{fig5}
\end{figure}

{\it Topological phase transition.} An important feature of the
topological SC phase with the Majorana zero bias peak is
the transport signature of phase transitions. Fig.~\ref{fig1}, if
continued for higher $V$, would display the interaction-induced transition
into a DW phase, where all previous main features such as
the Majorana peak and the proximity peaks disappear. Fig.~\ref{fig3}, if continued to
higher $\mu$, would eventually illustrate the evolution of the transport signal into a
trivial SC phase which also exists in the non-interacting
case. There, a separate investigation of the Majorana peak and the proximity
peaks, however, is quite hard to pursue because of the overpopulated
electron-like Bogoliubov band. 
On fundamental grounds of characterizing topological phase transitions,
the expectation is that at the transition between a trivial and a
topological SC phase, the bulk gap must close. In turn,
this implies that the Majorana peak cannot vanish without
the proximity peaks being affected as well. To illustrate this aspect and also to
choose a transition which might allow to draw connections to the
experimental setup where $\mu$ is held fixed~\cite{mourik-12s1222360},
we investigate the interaction-induced topological to trivial
SC transition at $\mu=0$ by varying $V$ from $-2$ to $-4$
(Fig.~\ref{fig5}). As we get closer to the transition,
the Majorana peak shrinks along with a successive vanishing of the
proximity gap until after the transition at $V_c~\sim -3.0$, the
proximity gap reopens without the Majorana peak. The fact that this
feature is well kept by the spectral function calculations in our DMRG
approach suggests that this behavior should generically be
observed for a topological SC phase transition in the transport signal
of Majorana wires.

{\it Summary and outlook.} We have shown that the Chebyshev expansion
method in DMRG allows us to obtain a detailed phase diagram of the
Kitaev chain in the presence of interactions via spectral
function calculations down to zero frequency. In the topological
SC phase we find a clean Majorana zero-bias peak. Investigating the
dependence of the spectral function on system parameters in the
presence of interactions, we find that while
$\mu$ changes the occupation of the hole-like versus the electron-like
Bogoliubov band, the Majorana zero-bias peak is hardly affected. The interactions modify the charge gap
and as such, for one effect, renormalize $\mu_c$ separating
the topologically trivial from the non-trivial SC phase in the wire. The
interactions affect the height-width ratio of the Majorana peak. As the interactions
reduce the bulk gap in the wire, the Majorana peak broadens and
vanishes along with the proximity gap peaks. We have investigated
differently tuned topological phase transitions and find that the bulk gap
closure manifests itself as a joint decay of the Majorana peak and the
proximity gap. Our analysis establishes
a starting point to endeavor the spinful Majorana wire models as well
as to study joint effects of disorder, temperature, and interactions
to establish a quantitative comparison with experimental
signatures. Including explicit estimates for transmission curves, it
will also be interesting to further analyze the possible renormalization of
AC and DC
conductance~\cite{safi-95prb17040,ponomarenko-95prb8666,maslov-95prb5539,thomale-11prb115330} in interacting Majorana wires.

\begin{acknowledgements}
We thank A. Akhmerov, B. Bauer, D. Goldhaber-Gordon, L. P. Kouwenhoven, F. Pollmann, and D.
Schuricht for helpful discussions. R.T.
is supported by DFG-SPP 1458. 
S.R. is supported by DFG-FOR 960 and DFG-SPP 1666.
\end{acknowledgements}


\end{document}